\documentclass[12pt]{iopart}
\usepackage{iopams}
\usepackage{graphicx}

\usepackage{citesort}

\begin{document}

\title[Terahertz Science and Technology of Carbon Nanomaterials]{Terahertz Science and Technology of Carbon Nanomaterials}

\author{R. R. Hartmann$^1$, J. Kono$^2$, and M. E. Portnoi$^3$}

\address{
$^1$ Physics Department, De La Salle University, 2401 Taft Avenue, Manila, Philippines
}

\address{
$^2$ Department of Electrical and Computer Engineering and Department of Physics and Astronomy, Rice University, 6100 Main St., MS-378, Houston, Texas 77005, USA
}

\address{
$^3$ School of Physics, University of Exeter, Stocker Road, Exeter EX4 4QL, UK
}

\ead{
\mailto{Richard.Hartmann@dlsu.edu.ph},
\mailto{kono@rice.edu},
\mailto{M.E.Portnoi@exeter.ac.uk}
}

\begin{abstract}
The diverse applications of terahertz radiation and its importance to fundamental science makes finding ways to generate, manipulate, and detect terahertz radiation one of the key areas of modern applied physics. One approach is to utilize carbon nanomaterials, in particular, single-wall carbon nanotubes and graphene. Their novel optical and electronic properties offer much promise to the field of terahertz science and technology. This article describes the past, current, and future of the terahertz science and technology of carbon nanotubes and graphene. We will review fundamental studies such as terahertz dynamic conductivity, terahertz nonlinearities, and ultrafast carrier dynamics as well as terahertz applications such as terahertz sources, detectors, modulators, antennas, and polarizers.
\end{abstract}


\maketitle

\section{Introduction}
The last three decades have witnessed explosive progress in the research and technology of carbon-based nanostructures highlighted by the discovery of fullerenes \cite{KrotoetAl85N}, rediscovery in the 1980s of the detonation technique of producing nanodiamonds \cite{GreineretAl88N} which resulted in a dramatic increase in the studies of their properties and applications \cite{MochalinetAl12NN}, syntheses of carbon nanotubes \cite{Iijima91N} and the unprecedented success of a simple scotch tape exfoliation technique \cite{NovoselovetAl04S} leading to the spectacular rise of graphene \cite{GeimNovoselov07NM}. The same period has seen wide-range efforts in bridging the so-called terahertz (THz) gap.

THz radiation lies between its better studied counterparts: microwave and infrared radiation in the electromagnetic spectrum. In this frequency range, electronic transport and optical phenomena merge with one another, and classical waves (in the microwave region) make the transition to quantum mechanical photons (in the optical regime). Therefore, understanding THz phenomena requires a multi-perspective approach.  In condensed matter, many elementary and collective low energy excitations occur in the THz range such as plasmons, magnons, and superconducting energy gaps, and the majority of dynamical phenomena in solids such as scattering, tunneling, and recombination occur on characteristic times scales of picoseconds, i.e., in the THz frequency range.

Filling the THz gap is a challenging area in modern device physics \cite{LeeWankeAl07S}, for this region of the electromagnetic spectrum presents difficulties in both generating coherent sources and creating sensitive detectors. However, the rewards of exploiting this gap are great, owing to the diverse applications of THz radiation. For example, the vibrational breathing modes of many large molecules occur in the THz domain making THz spectroscopy a potentially powerful tool for the identification and characterization of biomolecules \cite{THz_15,THz_16,THz_17,THz18,THz_19}. Furthermore, the non-ionizing nature of THz radiation means that it is seen by many as the future of imaging technology, and it also has promising applications in biomedicine and biosensing. As well as utilizing THz technology for pharmaceutical research \cite{THz_51} and biomedical diagnostic devices, other potential uses range from security applications such as the sensing and detection of biological hazards and explosives \cite{LeeWankeAl07S}, through to communication technology and astrophysics.


In the last decade, there has been significant progress made in the development of ultrafast laser based THz sources \cite{THz_4}, quantum cascade \cite{THz_5} and free electron lasers \cite{THz_6}, which operate in the THz range, as well as synchrotron-based THz sources \cite{THz_7}. However, current THz sources and detectors often suffer from low output power, are often considerable in size, and operate at liquid helium temperatures. For this reason, the search for ultra bright, coherent, and compact THz sources and detectors is one of the key areas of modern applied physics \cite{THz_8}. The unique position of the THz range, in the gap between the parts of electromagnetic spectrum that are accessible by either electronic or optical devices, leads to an unprecedented diversity in approaches to bridging this gap \cite{FergusonZhang02NM,Avrutin_SPS_88,Kruglyak_TPL_05,Mikhailov07EPL,DragomanDragoman04PQE,LeeWankeAl07S,Tonouchi07NP,ShumyatskyAlfano11JBO,UlbrichtetAl11RMP,Nagatsuma11IEE}. One approach is to utilize carbon nanomaterials, in particular, graphene and its one dimensional counterpart, the carbon nanotube \cite{DragomanDragoman04PQE}, whose optical and electronic properties are as interesting and diverse as their potential applications.

\subsection{What are carbon nanomaterials?}

\begin{figure}
\begin{center}
\includegraphics*[height=40mm]{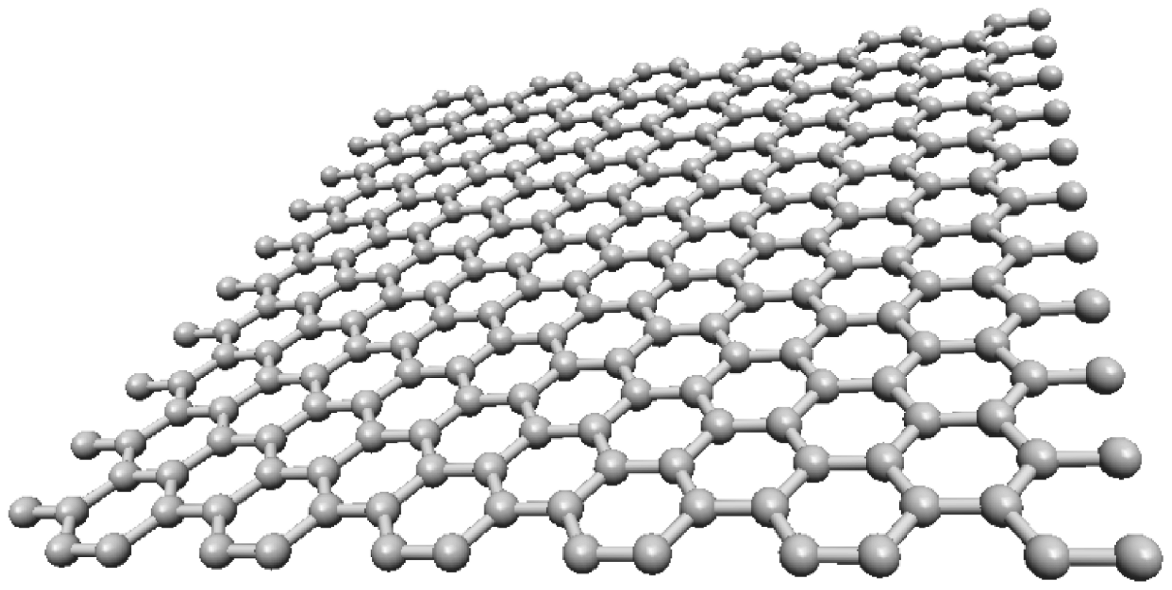}\\
\includegraphics*[height=40mm]{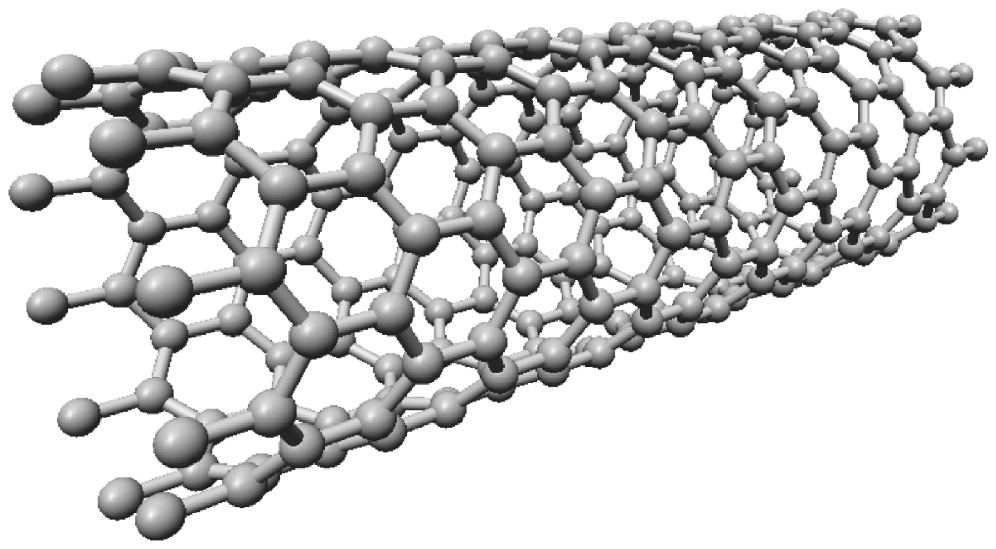}
\includegraphics*[height=40mm]{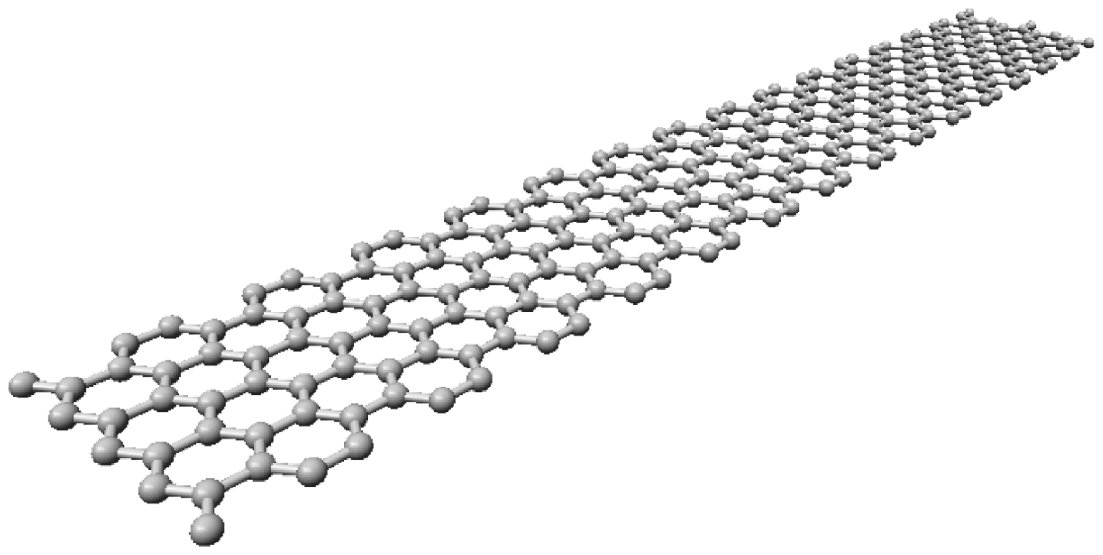}
\end{center}
\caption{
Graphene (top) is formed by carbon atoms arranged in a two-dimensional honeycomb crystal lattice. Graphene can be rolled into a cylinder to form carbon nanotubes (bottom left) or cut into graphene nanoribbons (bottom right).
}
\label{fig:nanomat}
\end{figure}

Carbon's ability to exist in many different forms is due to the fact that carbon's four valence electrons may hybridize in many ways. This hybridization may be $\mathrm{sp}$, $\mathrm{sp}^{2}$ or $\mathrm{sp}^{3}$, allowing carbon to form linear chains, planar sheets and tetrahedral structures. $\mathrm{sp}^{2}$-bonded carbon can form a honeycomb crystal lattice, one atom thick, known as graphene. This one atom thick sheet of carbon is the building block of many carbon nanomaterials (see figure~\ref{fig:nanomat}). Graphene can be rolled into a seamless cylinder to form carbon nanotubes, cut into graphene nanoribbons or fashioned into buckyballs. These low dimensional forms of nanocarbon could serve as the building blocks of carbon-based optoelectronic devices of the future.


\subsubsection{Graphene}
Graphene's carbon atom's $2s$, $2p_{x}$ and $2p_{y}$ orbitals form $\sigma$ bonds to their nearest neighbors, and these bonds determine the crystal's structural properties. The remaining $2p_{z}$ orbitals hybridize to form weaker, more delocalized $\pi$ bonds, and these dictate the optical and transport properties of the material. In general, the electronic structure of carbon nanomaterials can be described using a simple-tight binding model \cite{SaitoDresselhaus98Book} of graphene's $\pi$-electrons.
Unlike conventional systems, whose charge carriers are described by the Schr\"{o}dinger equation with an effective mass, graphene's charge carriers are described by the same equation used to describe two-dimensional massless Dirac fermions, the Dirac-Weyl equation \cite{Wallace_Phys_Rev_47,Novoselov05Nat,CastroNeto_Rmp_09}. The consequence of this unusual spectrum is that many effects that were once the in realms of high-energy particle physics are now measurable in a solid state system \cite{Katsnelson_NatPhys_06,Katsnelson_SSC_07}. Graphene exhibits unusual mechanical, thermal and, most of all, electronic properties \cite{CastroNeto_Rmp_09}, which allows the observation of interesting effects such as the suppression of backscattering \cite{AndoetAl98JPSJ}, an unconventional quantum Hall effect \cite{Novoselov05Nat,Zhang05Nat}, and a range of effects based on the Klein paradox \cite{Katsnelson_NatPhys_06}. The existence of massless Dirac fermions has been confirmed by magnetotransport measurements such as the integer quantum Hall effect \cite{Novoselov05Nat,Zhang05Nat}, and the linear spectrum results in an optical conductance defined by the fine structure constant \cite{MaketAl08PRL,NairetAl08Science,AndoetAl02JPSJ}.
The relativistic nature of graphene's charge carriers makes confinement difficult owing to the fact that particles do not experience back scattering in a smooth electrostatic potential \cite{AndoetAl98JPSJ,Katsnelson_NatPhys_06}. However, lateral confinement can be achieved by cutting graphene into nanoribbons, therefore allowing graphene to be used as a switching device. The geometrical confinement effects in graphene nanoribbons are similar to carbon nanotubes, but the boundary conditions imposed on the wavefunction due to edge type can lead to drastically different electronic properties \cite{Nakada96PRB,Zheng07PRB,Brey06PRB} such as the existence of edge states.

\subsubsection{Carbon Nanotubes}

\begin{figure}
\begin{center}
\includegraphics*[height=60mm]{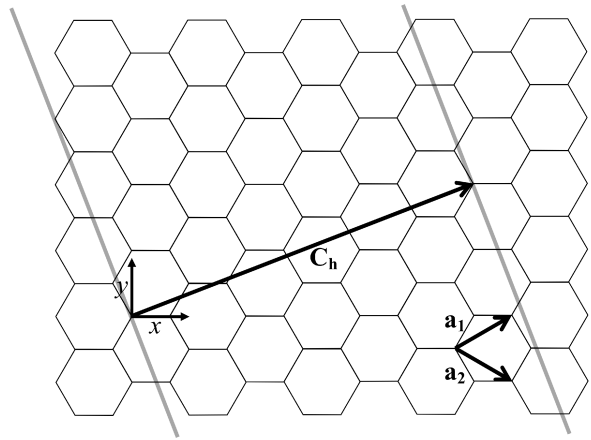}
\end{center}
\caption{
An unrolled carbon nanotube. The cylinder is formed by rolling the graphene sheet along the chiral vector, $\bf{C_{h}}$, such that its start and end points coincide. $\bf{a_{1}}$ and $\bf{a_{2}}$ are the graphene primitive lattice vectors.
}
\label{fig:CNT}
\end{figure}

A single-walled nanotube (SWCNT) is formed by rolling a graphene sheet into a seamless cylinder \cite{SaitoDresselhaus98Book,DresselhausetAl01Book,ReichBook04,revAnantram06}. The way in which the nanotube is rolled is described by the chiral vector $\mathbf{C_{h}}$, which connects crystallographically equivalent sites of the unrolled graphene lattice (see figure~\ref{fig:CNT}). The chiral vector, is defined via the graphene primitive lattice vectors, $\mathbf{a_{1}}$ and $\mathbf{a_{2}}$, as $\mathbf{C_{h}}\left(n,m\right)=n\mathbf{a_{1}}+m\mathbf{a_{2}}\equiv\left(n,m\right)$, where $n$ and $m$ are integers. If $n=m$ the nanotubes are called armchair nanotubes, and when $m=0$ the nanotubes are referred to as zigzag nanotubes; all other cases are classified as chiral nanotubes. For the case of $n-m=3p$, where $p$ is a non-zero integer, the tubes are known as quasi-metallic tubes \cite{Hamadaetal92PRL}.

The manner in which the tube is rolled strongly dictates its electronic properties: the SWCNT can either be metallic, a narrow-gap semiconductor, or semiconducting. Within the frame of a simple zone-folding model of the $\pi$-electron graphene spectrum, armchair and quasi-metallic carbon nanotubes can be considered as one-dimensional analogs of graphene, since in this model the electron low-energy spectrum is linearly dependent on the electron wavevector. The electron energy spectrum, $\varepsilon\left(k\right)$, of the aforementioned SWCNTs is given by $\varepsilon\left(k\right)=\pm\hbar v_{\mathrm{F}}\left|k\right|$, where $k$  is measured from where the conduction and valence bands coincide and $v_{\mathrm{F}}\approx9.8\times10^{5}$ m/s is the Fermi velocity of graphene. However, due to curvature effects quasi-metallic SWCNTs are in fact narrow-gap semiconductors \cite{Hamadaetal92PRL}, their bandgap given by
$\varepsilon_g={\hbar v_F a_{\mbox{\scriptsize{C-C}}} \cos
3\theta}/(8 R^2)$~\cite{KaneMele97PRL,Gunlycke06}, where
$a_{\mbox{\scriptsize{C-C}}}=1.42\;$\r{A} is the nearest-neighbor
distance between two carbon atoms, $R$ is the nanotube radius, and
$\theta=\arctan [\sqrt{3}m/(2n+m)]$ is the chiral
angle~\cite{SaitoDresselhaus98Book} and their band gaps can be of the order of THz.

\subsection{Why are they good for THz science and technology?}
While initial investigations on these materials concentrated on DC characteristics, recent theoretical studies have instigated a flurry of new experimental activities to uncover unusual AC properties.   Both nanotubes and graphene are expected to show exotic THz dynamics that can lead to innovative optoelectronic applications~\cite{PortnoietAl06SPIE,PortnoietAl08SM,Mikhailov09MJ,RenetAl12JIMT,OtsujietAl12JPD}.  These properties are inherently related to their unique, low-dimensional band structure, combined with many-body interactions of quantum-confined carriers. In the presence of external magnetic fields and electric fields, certain types of nanotube develop strong THz optical transitions, giving rise to the possibility of utilizing them as highly tunable, optically-active materials in THz devices \cite{KibisPRB05,PortnoietAl08SM,PortnoietAl09IJMPB,Batrakov_Physica_B_2010,NemilentsauetAl07PRL}, and their desirable electronic properties and highly anisotropic optical properties make them ideally suited for THz antenna and polarizer applications \cite{JeonetAl02APL,JeonetAl04JAP,AkimaetAl06AM,RenetAl09NL,KyoungetAl11NL,RenetAl12NL,Ren13PRB}. Nanotubes also hold the promise of ballistic THz transistors~\cite{Burke04SSE} that could supersede traditional silicon technology.

As a gapless semiconductor with ultra high carrier mobility, graphene is a natural material for THz applications, and the ability to modify graphene's charge carrier density through electronic gating makes graphene ideally suited for optoelectronic applications. Its charge carriers can be further manipulated by externally applied magnetic fields, allowing tunable broadband detectors to be realized \cite{kawano2013wide}. Graphene is also predicted to be a gain medium for THz lasers \cite{RyzhiietAl07JAP} which can operate at room temperature, an exciting prospect considering the dearth of THz sources. Equally exciting is graphene's ability to support surface plasmon modes in the THz regime ~\cite{Grigorenko12NatPhoton,Gao13NanoLetts}, which can be tuned by electronic gating, therefore paving the way for novel graphene-based plasmonic devices that operate in the THz regime.


\section{THz Science and Technology of Carbon Nanotubes}

\subsection{Dynamic Conductivity}

For one-dimensional (1-D) electron systems, including SWCNTs, there have been theoretical calculations of $\sigma(\omega)$, taking into account interactions and disorder to varying degrees (see, e.g., \cite{SablikovShchamkhalova97JETP,RoschAndrei00PRL,Ando02JPSJ,Burke02IEEE,PustilniketAl06PRL,NakanishiAndo09JPSJ} and pp.~219-237 of \cite{Giamarchi04Book}).   Specifically for a metallic SWCNT, Ando~\cite{Ando02JPSJ} calculated $\sigma(\omega)$ within a self-consistent Born approximation, which indicated that there can exist non-Drude-like conductivity, depending on the range of scattering potentials.

In a 1-D conductor, electron-electron interactions lead to a breakdown of the Fermi liquid model. In this instance, the charge carriers are described as a Tomonaga-Luttinger liquid \cite{Tomonaga50PTP,Luttinger63JMP}. Tomonaga-Luttinger liquids were studied both theoretically \cite{Egger97PRL,Kane07PRL5086} and observed experimentally in single \cite{Bockrath99Nat,Yao99Nat,Postma00PRB,Eggert04PRL} and multi-wall nanotubes \cite{Tarkiainen01PRB,Bachtold01PRL}. Burke showed that these modes 1-D plasmons can be excited at gigahertz frequencies \cite{Burke02IEEETTNT129} thus demonstrating the feasibility of THz nanotube transistors, amplifiers and oscillators.

A number of experimental THz/far-infrared spectroscopic studies have been performed over the last decade on SWCNTs of various forms~\cite{BommelietAl96SSC,UgawaetAl99PRB,ItkisetAl02NL,JeonetAl02APL,JeonetAl04JAP,JeonetAl05JAP,AkimaetAl06AM,BorondicsetAl06PRB,NishimuraetAl07APL,KampfrathetAl08PRL,RenetAl09NL,SlepyanetAl10PRB,RenetAl12NL,Ren13PRB}, producing an array of conflicting results with contradicting interpretations.  This is partly due to the widely differing types of samples used in these studies -- grown by different methods (HiPco, CoMoCAT, CVD, Arc Discharge, and Laser Ablation) and put in a variety of polymer films that are transparent in the THz range.  Nanotubes in most of these samples were bundled and typically consisted of a mixture of semiconducting and metallic nanotubes with a wide distribution of diameters.  One common spectral feature that many groups have detected is a broad absorption peak around 4~THz (or 135~cm$^{-1}$ or $\sim$17~meV).  This feature, first observed by Ugawa {\it et al}.~\cite{UgawaetAl99PRB}, has been interpreted as interband absorption in quasi-metallic 
nanotubes with curvature-induced gaps~\cite{UgawaetAl99PRB,BorondicsetAl06PRB,NishimuraetAl07APL,KampfrathetAl08PRL} or absorption due to plasmon oscillations along the tube axis~\cite{JeonetAl02APL,SlepyanetAl10PRB,ShubaetAl12PRB,AkimaetAl06AM,NakanishiAndo09JPSJ,Ren13PRB,Zhang13NanoLetts}, but a consensus has not been achieved~\cite{SlepyanetAl10PRB}.


\subsection{THz Emitters and Detectors}
There are several promising proposals of using carbon nanotubes for THz applications: a nanoklystron utilizing extremely efficient high-field electron emission from nanotubes \cite{DragomanDragoman04PQE,ManoharaetAl05JVSTB,PortnoietAl06SPIE}, devices based on negative differential conductivity in large-diameter semiconducting nanotubes \cite{MaksimenkoSlepyan00PRL,PenningtonGoldsman03PRB}, high-frequency resonant-tunneling diodes \cite{DragomanDragoman04PE} and Schottky diodes \cite{Odintsov00PRL,Leonard00,YangetAl05APL,LuetAl06APL}, as well as frequency multipliers~\cite{Slepyan99,Slepyan01}, THz amplifiers~\cite{Dragoman05}, switches~\cite{Dragoman06} and antennas~\cite{Slepyan06}. THz detectors based upon antennas coupled to bundles \cite{Fu08APL,yngvesson2008experimental,carrion2009new,carrion2010single} and individual \cite{carrion2010single,chudow2012terahertz} metallic SWCNTs have also been demonstrated.

\begin{figure}
\begin{center}
\includegraphics*[width=0.6\textwidth]{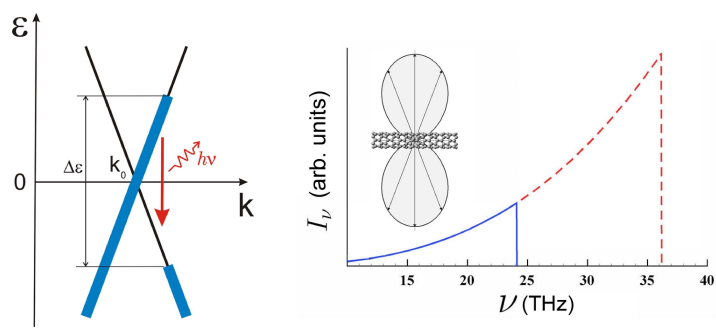}
\end{center}
\caption{
(a) The scheme of THz photon generation by hot carriers in
quasi-metallic SWNTs. (b) The spectral density of spontaneous
emission, $I_{\nu}$, as a function of frequency, $\nu$, for two values of applied
voltage: solid line for $V=0.1\;$V; dashed line for $V=0.15\;$V. Adapted from \cite{PortnoietAl08SM}.
}
\label{fig:inversion}
\end{figure}

A variety of proposals exist for using carbon nanotubes for THz emitters and detectors~\cite{KibisPRB05,PortnoietAl06SPIE,KibisetAl07NL,NemilentsauetAl07PRL,PortnoietAl08SM,PortnoietAl09IJMPB}. Kibis {\it et al}.~\cite{KibisetAl07NL} demonstrated that quasi-metallic carbon nanotubes emit THz radiation when a potential difference is applied across their ends. The electric-field induced heating of the electron gas results in a population inversion of optically active states with an energy difference within the THz range (see figure~\ref{fig:inversion}). In the ballistic regime, the spontaneous emission spectra of all quasi-metallic have a universal dependence on the photon frequency~\cite{Hartmann11PhD}, with the maximum of the spectral density of emission being controlled by the size of the applied voltage, which raises the possibility of utilizing this effect for high-frequency nanoelectronic devices. In figure~\ref{fig:inversion} the spectral density is shown for two values of  applied voltage.

It has been shown that chiral carbon nanotubes can be used as the basis of tunable frequency multipliers~\cite{KibisPRB05,PortnoietAl08SM}. An electric field applied normal to the nanotube axis gives rise to regions of the energy spectrum with negative effective-mass~\cite{KibisPRB05,PortnoietAl08SM}, which are accessible in moderate electric fields. The effect of the negative effective mass also leads to an efficient frequency multiplication in the THz range, which can be controlled by the applied electric field.

\begin{figure}
\begin{center}
\includegraphics*[width=0.9\textwidth]{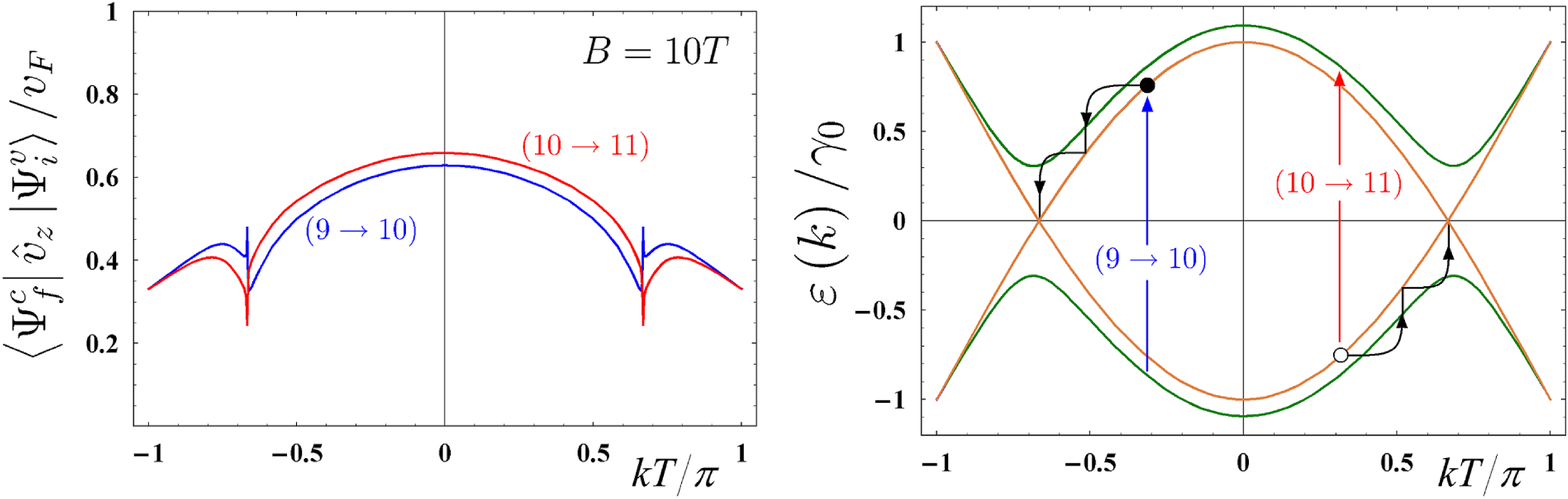}
\end{center}
\caption{A scheme for creating a population inversion between the
lowest conduction subband and the top valence subband of an armchair
SWNT in a magnetic field. The left plot shows the calculated matrix
elements of the relevant dipole optical transitions polarized
normally to the axis of a $(10,10)$ SWNT. The right plot shows
several energy subbands closest to the Fermi level and illustrates
the creation of photoexcited carriers and their non-radiative
thermalization. Adapted from \cite{PortnoietAl08SM}}
\label{fig:MEP}
\end{figure}

Armchair SWCNTs are truly gapless, but by applying a magnetic field along the nanotube a band gap can be opened~\cite{AjikiAndo93JPSJ,ReichBook04,KonoRoche06CRC,KonoetAl07Book}. For a $(10,10)$ SWNT in a field  of 10~T the band gap corresponds to approximately 1.6~THz. It transpires that the same magnetic field which opens the band gap also allows dipole optical transitions between the top valence subband and the lowest conduction subband~\cite{PortnoietAl08SM,PortnoietAl09IJMPB} which would otherwise be forbidden in the absence of a magnetic field~\cite{MilosevicetAl03PRB,JiangetAl04Carbon}. In the presence of the field, the van Hove singularity in the reduced density of states leads to a very sharp absorption maximum near the band edge, which results in a very high sensitivity in the photocurrent to the photon frequency, whose peak frequency is tunable by the size of the applied magnetic field. This scheme can also be used as a tunable emitter with a very narrow emission line. A population inversion can be achieved by optical pumping with the light polarized normally to the nanotube axis, as shown in figure~\ref{fig:MEP}.

In the absence of curvature, at $k=0$, optical transitions between the top valence subband and the bottom conduction subband in quasi-metallic SWCNTs are strictly forbidden by symmetry within the simple zone-folding model of the $\pi$-electron graphene spectrum. However, for zigzag quasi-metallic SWCNTs, dipole optical transitions are indeed allowed due to the gap opened in their energy spectrum by intrinsic curvature, which is of the order of THz~\cite{Hartmann11PhD,Haroz13Nanoscale}. As with the armchair nanotube, a magnetic field applied along the nanotube can be used to modify the optical selection rules. However in quasi metallic tubes, the magnetic field creates two different size band gaps, and therefore two peaks in the absorption spectra. Arrays of armchair and quasi-metallic SWCNTs could be used as the building blocks of THz radiation detectors, which would have a high sensitivity in the photocurrent to photon frequency. Furthermore, since the band gap of such SWCNTs can be controlled by the size of the applied magnetic field, such devices can be tunable.

Many-body (excitonic) effects, which dominate the optical properties of semiconducting SWCNTs~\cite{Ando97JPSJ,WangetAl05Science,MaultzschetAl05PRB2,ShaverKono07LPR}, are also important in narrow-gap SWCNTs. However, due to the quasi-relativistic character of the free-particle dispersion near the band edge of the narrow-gap SWCNTs, there is a spectacular decrease in the exciton binding energy \cite{Hartmann11PRB}. The binding energy scales with the band gap and therefore, excitonic effects should not dominate in narrow-gap nanotubes. Hence, the proposed THz applications of quasi-metallic nanotubes should be feasible.

Nemilentsau {\it et al}.~\cite{NemilentsauetAl07PRL} theoretically studied thermal radiation from an isolated finite-length carbon nanotube both in near- and far-fields.  The formation of the discrete spectrum in metallic nanotubes in the THz range is demonstrated due to the reflection of strongly slowed-down surface-plasmon modes from nanotube ends.

\subsection{THz Antennae and Polarizers}
Nanotubes' desirable electronic properties and highly anisotropic optical properties make them ideally suited for antenna and polarizer applications in the THz range. Both single   \cite{Hanson05IEEETAP,Hanson06IEEETAP,BurkeetAl06IEEE,Slepyan06,Slepyan08ICMMET,HaoIEEETNano,Wang08COL,Yue08CPB,Fichtner08ARC,Jornet10IEEEPEC}
and multi-wall  \cite{Hanson08IEEEAAPM,Shuba09PRB,Berres11IEEET} carbon nanotube antennae operating in the THz regime have been studied extensively, and bundles/arrays of nanotubes \cite{Hao06PRB,Shuba07PRB,Wang08IJIMW,Huang08IEEETNano,Maksimenko08PE,Ren13PRB} have been shown to outperform tubes in isolation, demonstrating a far superior antenna efficiency \cite{Ren13PRB}. The concept of a thermal nanoantenna was proposed by Nemilentsau {\it et al}.~\cite{NemilentsauetAl07PRL}.

Polarization anisotropy in the THz range was first noted by Jeon and co-workers by using partially aligned nanotube films~\cite{JeonetAl02APL,JeonetAl04JAP}.  Akima {\it et al}.~\cite{AkimaetAl06AM} also demonstrated anisotropy. More recently, using extremely well aligned ultralong carbon nanotubes, Ren and co-workers demonstrated that carbon nanotubes can be perfect THz polarizers~\cite{RenetAl09NL}. Kyoung and co-workers used aligned multi-wall carbon nanotubes to demonstrate similarly strong anisotropy~\cite{KyoungetAl11NL}.

\begin{figure}
\begin{center}
\includegraphics*[height=45mm]{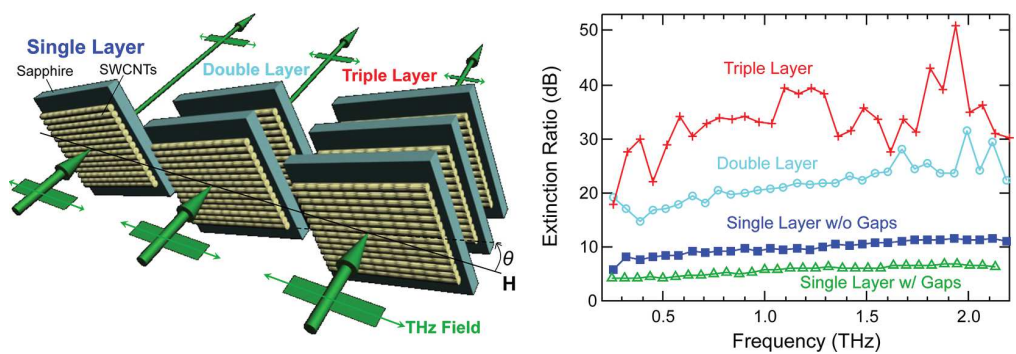}
\end{center}
\caption{
Scheme showing the use of multiple SWCNTs films to produce high performance polarizers. The extinction ratio of the THz polarizers with different thicknesses is shown as a function of frequency in the $\sim0.4-2.2$ THz range. Adapted from \cite{RenetAl12NL}.
}
\label{Polarizer}
\end{figure}

More recently, Ren {\it et al}. showed that adding additional layers of films (as shown in figure \ref{Polarizer}) will enhance device performance considerably, achieving ideal broadband THz properties in a triple-stacked film: 99.9\% degree of polarization and extinction ratios of 10$^{-3}$ (or 30\,dB) from $\sim$0.4 to 2.2~THz~\cite{RenetAl12NL}, which is two orders of magnitude higher than that of a single layer. Not only do these multi-layer devices outperform wire-grid polarizers in the THz range, but since they are based on the intrinsic properties of the tubes (rather than geometric effects like wire-grid polarizers), such devices can operate beyond the THz range. Despite the macroscopic thickness of the layers, for polarizations normal to the nanotube axis there is almost zero absorption. However, in contrast to the Drude model, they show that when the polarization is parallel to the nanotube axis the transmittance decreases with increasing frequency. This is attributed to the existence of a peak in THz conductivity, which was discussed in section 2.1.

\subsection{THz Transistors}
\begin{figure}
\begin{center}
\includegraphics*[height=50mm]{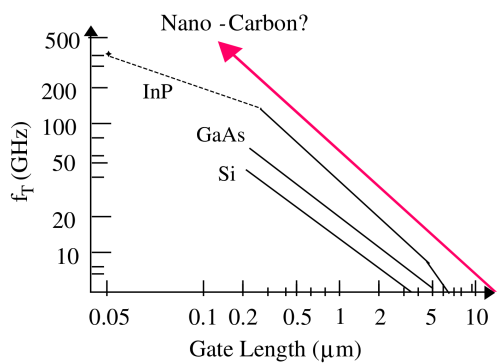}
\end{center}
\caption{
Transistor cutoff frequency, $f_{\mathrm{T}}$, versus gate length. References are: solid lines~\cite{sze2006physics}; dashed
line~\cite{endoh2000high}; nano-carbon prediction \cite{Burke04SSE}. From \cite{Burke04SSE}.
}
\label{fig:Burke}
\end{figure}
The demonstration of ballistic transport \cite{JaveyetAl03Nature} and linearity \cite{Baumgardner07APL} in carbon nanotube field-effect transistors demonstrates that nanotubes are possible candidates to replace traditional silicon technology. By analyzing the influence of quantum capacitance, kinetic inductance, and ballistic transport on the high-frequency properties of nanotube transistors, Burke predicted a cutoff frequency of 80\,GHz/$L$, where $L$ is the gate length in microns, opening up the possibility of a ballistic THz nanotube transistor~\cite{Burke04SSE}. Figure~\ref{fig:Burke} shows how Burke's prediction compares against other technologies. Several other theoretical studies have also shown that nanotube field effect transistors can potentially operate in the THz frequency regime \cite{Guo05IEEETN,Alam05APL,Hasan06IEEETN,Pulfrey09SSE,Kienle09PRL,koswatta2011ultimate}.

Experimentally, several field-effect transistors based upon individual nanotubes, operating in the low GHz range have been achieved  \cite{LietAl04NL,Wang07IEEENT,Chaste08NL,Louarn07APL,Yu06APL,Rosenblatt05APL,Pesetski06APL,Nougaret10APL,Cobas11IEEETMTT}. However intrinsic cutoff frequencies in the THz regime have yet to be realized. The performance of such devices have been limited by parasitic effects and impedance mismatches. One approach to increase the extrinsic cutoff frequency is to utilize networks or arrays of aligned nanotubes \cite{Kocabas08PNAS,Kocabas09NL,steiner2012high,Nougaret09APL,Wang11ACSN,Louarn07APL,Hu04NL,Kocabas07NL,Kang07NatN,Engel08ACSN,Chimot07APL}.
Most recently, Che {\it et al}. \cite{Che13ACSN} demonstrated that such an array of carbon nanotube transistors can achieve an extrinsic current-gain cutoff frequency of 25 GHz.

\subsection{Saturable absorbers for ultrafast photonics}
\begin{figure}
\begin{center}
\includegraphics*[width=90mm]{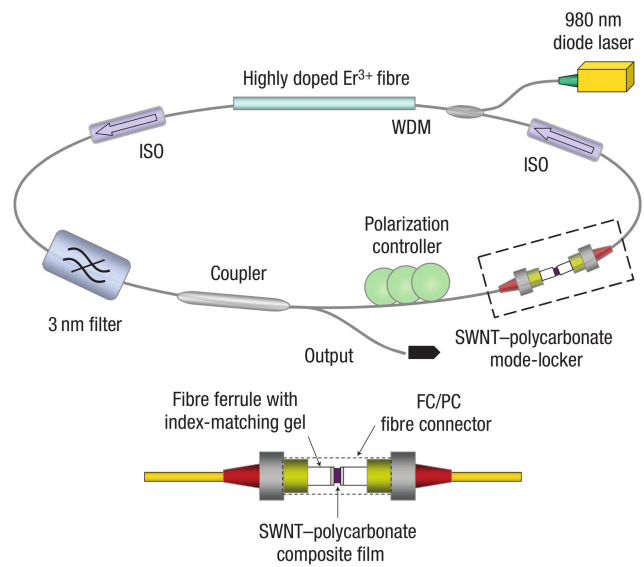}
\end{center}
\caption{
Ultrafast fiber laser setup and mode-locker assembly. This schematic shows the standard fibre-optic components such as an optical isolator (ISO), a wavelength division multiplexer (WDM), a power splitter and a polarization controller that are found in the ring cavity. The total length of the cavity is about 13.3 m. Adapted from \cite{wang2008wideband}.
}
\label{fig:ultrafast}
\end{figure}
One of the most common of the currently used methods of THz spectroscopy for both laboratory research \cite{UlbrichtetAl11RMP} and commercial sensors \cite{Tonouchi07NP} is based on shining sub-picosecond near-infrared light pulses onto a photoconductive emitter. The most expensive and bulky element of the THz spectroscopy setup is a short-pulsed laser source, which is used for both THz radiation generation and its detection after it passes through a sample and an electro-optical birefringent crystal \cite{THz_4}. The search for cheap, compact sub-picosecond laser sources for THz spectroscopy applications has been driving the development of short-pulsed semiconductor lasers for several decades \cite{avrutin2000monolithic}, with the majority of systems employing a mode-locking technique, in which a saturable absorber is used to convert the laser's continuous wave output into a train of ultrashort pulses. It has been demonstrated \cite{wang2008wideband,hasan2009nanotube} that single-walled carbon nanotubes and nanotube-polymer composites are excellent saturable absorbers because of their sub-picosecond recovery time, low saturation intensity and mechanical and environmental robustness. Carbon nanotube-based saturable absorbers are not only free from rigid recovery time limitations, which are unavoidable in conventional semiconductor systems \cite{Avrutin_SPS_88}, but also have an additional advantage of operating within a much broader range of frequencies. Figure~\ref{fig:ultrafast} shows a setup of a short-pulsed fiber-optics laser containing a nanotube-based saturable absorber.  Generation of 133 fs pulses with 33.5 mm spectral width and approximately 0.07 percent amplitude fluctuation was reported for a stretched-pulse fiber laser based on a nanotube saturable absorber \cite{sun2010ultrafast}. Notably, sub 200 fs pulse generation was also achieved from a mode-locked fiber laser with a saturable absorber based on graphene \cite{popa2010sub}.

\section{THz Science and Technology of Graphene}
\subsection{Dynamic Conductivity}
The AC dynamics of Dirac fermions in graphene have attracted much theoretical attention -- the influence of linear dispersions, two-dimensionality, and disorder has been extensively discussed by many theorists~\cite{ShonAndo98JPSJ,ZhengAndo02PRB,AndoetAl02JPSJ,PeresetAl06PRB,GusyninSharapov06PRB,GusyninetAl06PRL,GusyninetAl07PRL,RyuetAl07PRB,AbergelFalko07PRB,Mishchenko07PRL,GusyninetAl07PRB,FalkovskyVarlamov07EPJB,FalkovskyPershoguba07PRB,SheehySchmalian07PRL,HerbutetAl08PRL,KoshinoAndo08PRB,Mishchenko08EPL,LewkowiczRosenstein09PRL,IngenhovenetAl10PRB,JuricicetAl10PRB,VaskoetAl12PRB}.  However, the influence of electron-electron interactions on the optical conductivity of graphene is somewhat controversial.  Theoretical studies using different methods have led to different conclusions as to the magnitude of many-body corrections to the Drude-like intraband optical conductivity (see, e.g.,~\cite{Mishchenko07PRL,SheehySchmalian07PRL,HerbutetAl08PRL,Mishchenko08EPL} and references cited therein).

Experimentally, while a number of studies have confirmed the so-called universal optical conductivity $\sigma(\omega) = e^{2} /4 \hbar$ for {\em interband} transitions in a wide spectra range~\cite{NairetAl08Science,MaketAl08PRL,LietAl08NP}, successful experimental studies of the {\em intraband} conductivity have been reported only recently~\cite{DawlatyetAl08APL,ChoietAl09APL,HorngetAl11PRB,LiuetAl11JAP,YanetAl11ACS,RenetAl12NL2,MaengetAl12NL,Sensale-RodriguezetAl12NC}.  Also, two groups have used microscopy techniques to map out dynamic conductivity non-uniformity in the GHz~\cite{TalanovetAl10ACS} and THz~\cite{BuronetAl12NL} range and many successful dynamic conductivity measurements in magnetic fields (cyclotron resonance) exist~\cite{SadowskietAl06PRL,JiangetAl07PRL,DeaconetAl07PRB,HenriksenetAl08PRL,OrlitaetAl08PRL,NeugebaueretAl09PRL,HenriksenetAl10PRL,CrasseeetAl11NP,WitowskietAl10PRB,OrlitaetAl11PRL,CrasseeetAl11PRB,BooshehrietAl12PRB}.


\subsection{THz Gain and Lasing in Optically Excited Graphene}

\begin{figure}
\begin{center}
\includegraphics*[height=50mm]{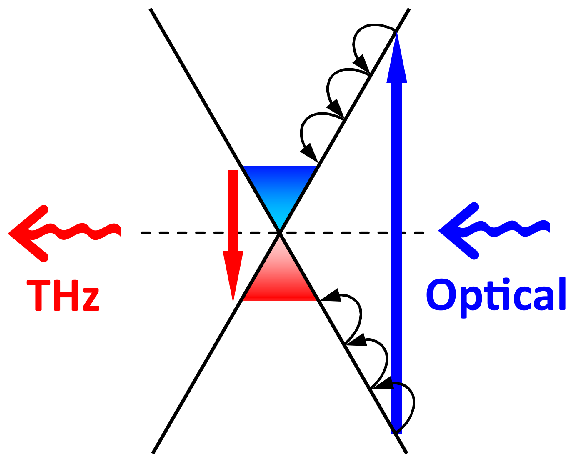}
\includegraphics*[height=50mm]{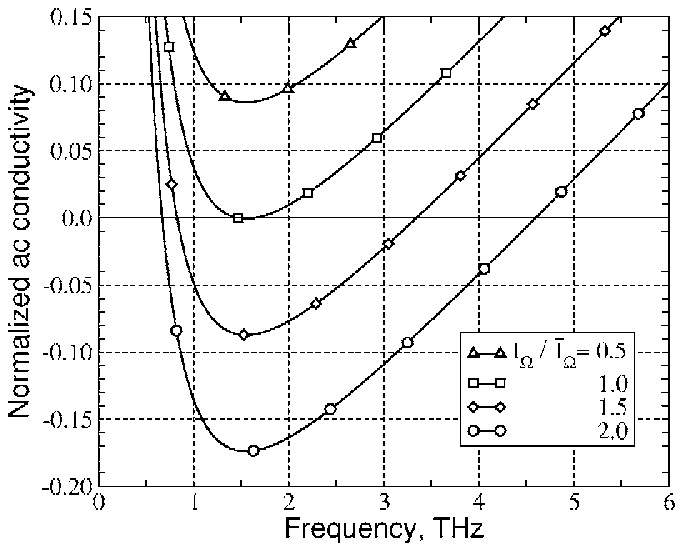}
\end{center}
\caption{
The left plot shows a scheme for creating a population inversion in graphene. Electrons and holes generated by optical pumping non-radiatively thermalize and aggregate towards the dirac points where they can recombine to emit THz radiation.
The right plot shows the frequency dependence of normalized ac conductivity at different intensities of optical radiation, $I_{\Omega}$, with the photon energies $\hbar\Omega$. $\overline{I}_{\Omega}$ is the threshold intensity of optical radiation. From~\cite{RyzhiietAl07JAP}.
}
\label{fig:Laser}
\end{figure}

\begin{figure}
\begin{center}
\includegraphics*[height=50mm]{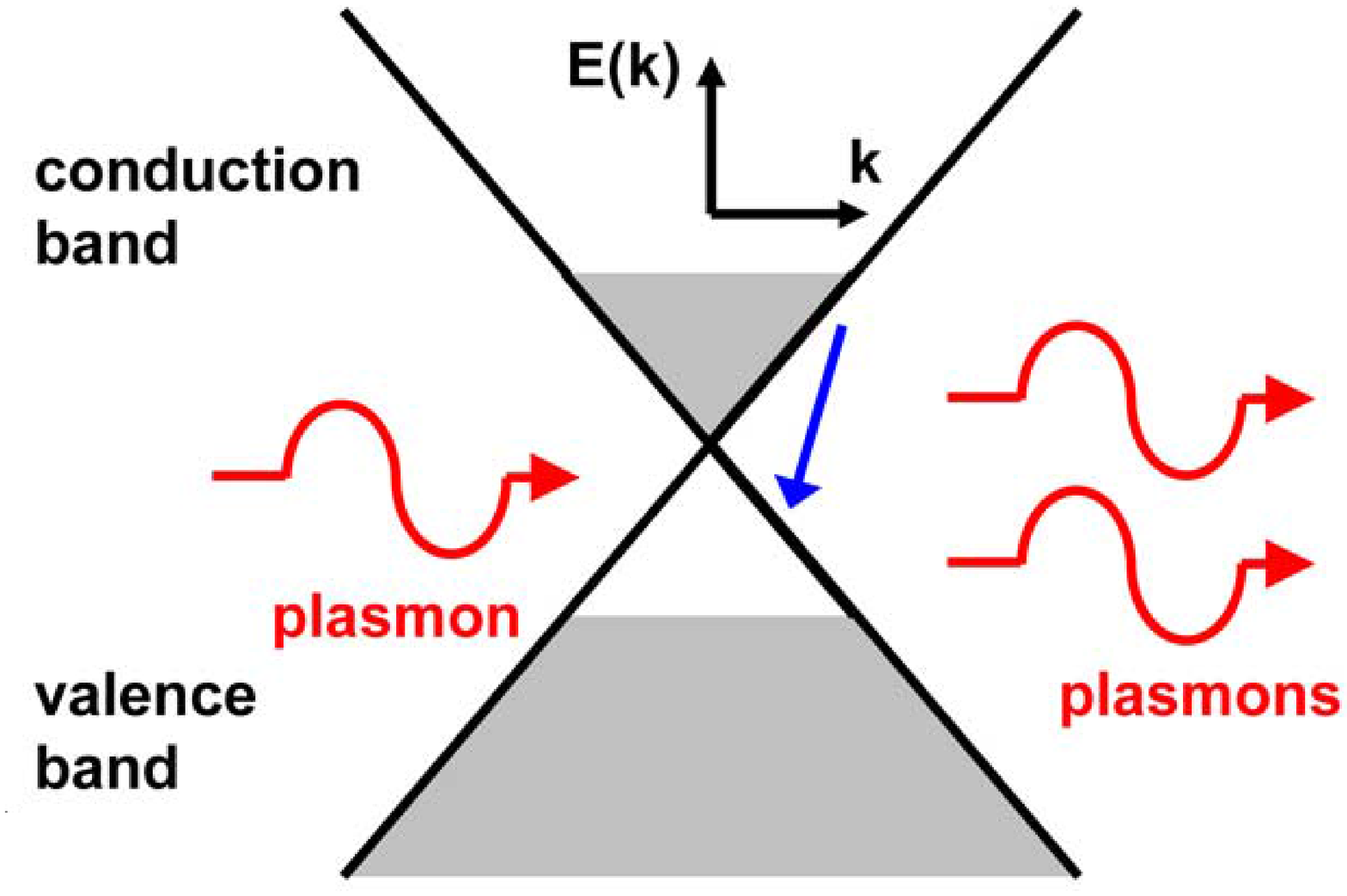}
\includegraphics*[height=50mm]{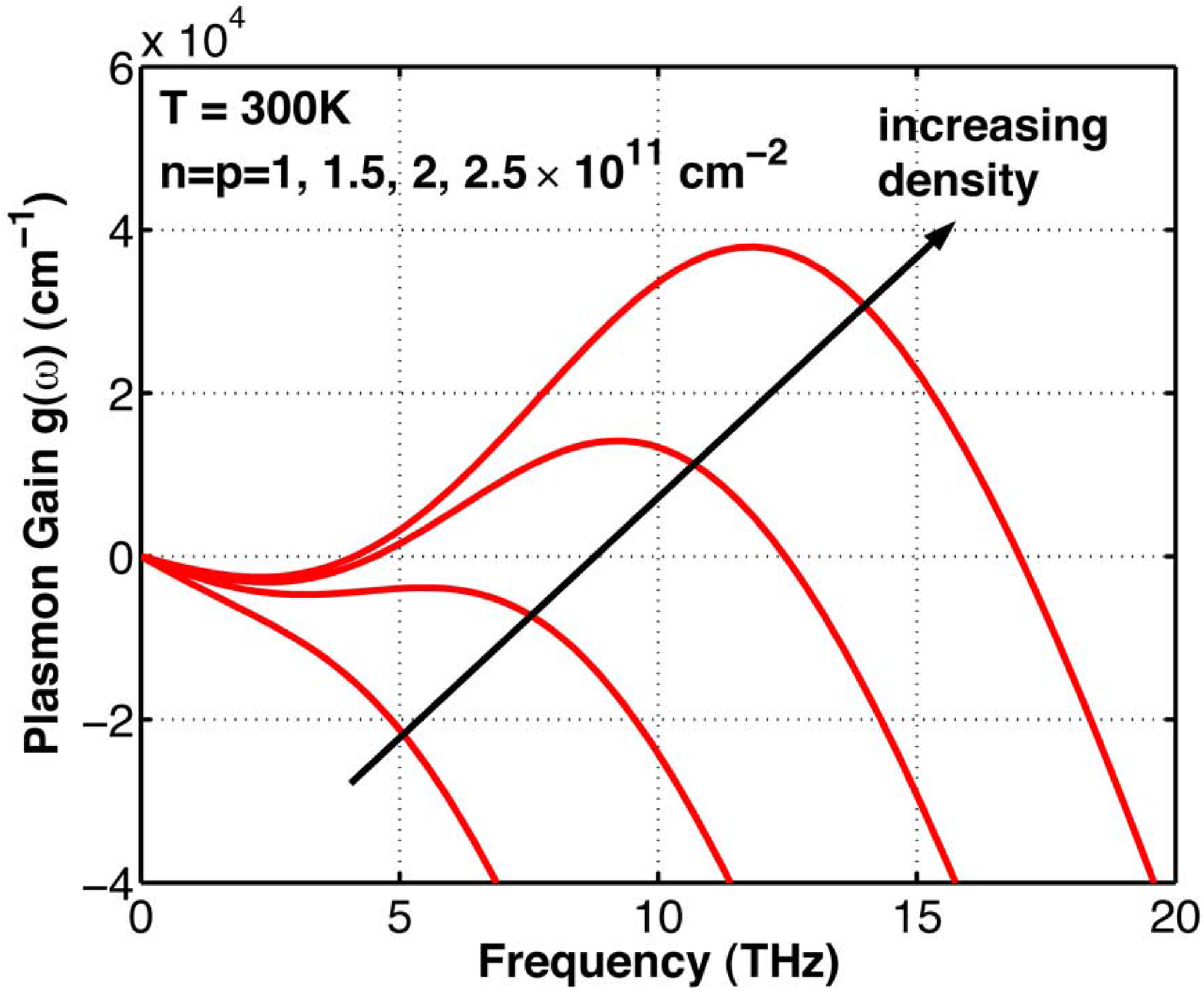}
\end{center}
\caption{
The left plot shows plasmon amplification through stimulated emission in population inverted graphene. The right plot shows the net plasmon gain (interband gain minus intraband loss) in graphene at $300$~K plotted for different electron-hole densities ($n = p = 1, 1.5, 2, 2.5 \times10^{11}\,\mathrm{cm}^{-2}$). The assumed values of the Fermi velocity and scattering time are $10^8$~cm/s and $0.5$~ps respectively. Adapted from~\cite{RanaIEEE08NanoT}.
}
\label{fig:Rana}
\end{figure}

An outstanding theoretical prediction is THz amplification in optically pumped graphene, proposed by Ryzhii and co-workers~\cite{RyzhiietAl07JAP,SatouetAl08PRB,DubinovetAl09APE,SatouetAl11JJAP,Satou13JAP}.  They demonstrated that sufficiently strong optical pumping will result in population inversion, making the real part of the net AC conductivity negative, i.e., amplification (see figure~\ref{fig:Laser}). Due to the gapless energy spectrum, this negative AC conductivity takes place in the range of THz frequencies. They calculated the dynamic conductivity of a non-equilibrium 2D electron-hole system in graphene under interband optical excitation. Both interband and intraband transitions were taken into account in their model. It has been shown that under population inversion it is possible to achieve plasmon amplification through stimulated emission \cite{Rana07arXiv,RanaIEEE08NanoT}. The predicted net plasmon gain for different electron-hole densities at $300$~K is shown in figure~\ref{fig:Rana}. Several plasmonic amplifiers and oscillations in graphene devices have been proposed for generating THz waves~\cite{RanaIEEE08NanoT,RyzhiietAl09APE,RyzhiietAl08JPCM,Dubinov11JAP,PopovetAl12PRB}.

It has been predicted that THz lasing can be realized at room temperature in optically pumped lasers utilizing Fabry-P{\'e}rot resonators \cite{DubinovetAl09APE,Ryzhii09JAP106} and dielectric waveguides \cite{RyzhiietAl10JAP}. It has also been shown that population inversion can be achieved by current injection in graphene layer and multilayer graphene heterostuctures \cite{Ryzhii07JAP46,Ryzhii11JAP110,Otsuji12MRSB,Ryzhii13arXiv}, and a novel voltage-tunable coherent THz emitters based upon multilayer graphene-boron-nitride heterostructures has been proposed by Mikhailov \cite{Mikhailov13PRB}. Freitag {\it et al}.~observed thermal infrared emission from biased graphene~\cite{FreitagetAl10NN}, and Ramakrishnan {\it et al}.~observed coherent THz emission from optically excited graphite~\cite{RamakrishnanetAl09OE}.

A number of groups have performed ultrafast pump-probe spectroscopy to study carrier dynamics \cite{Sun08PRL,Dawlaty08APL,Newson09OE,Breusing09PRL,Plochocka09PRB,wang10APL,Lui10PRL,Obraztsov11NL,Hale11PRB,Breusing11PRB,Sun11SSC,Brida13Nat,GeorgeetAl08NL,ChoietAl09APL,Karasawa11JITW,Otsuji11SPIE,Strait11NL,Boubanga-TombetetAl12PRB,DochertyetAl12NC,Tielrooij13NatP,PrechteletAl12NC,TanietAl12PRL,LietAl12PRL,Winnerl11PRL}, including optical-pump THz-probe experiments~\cite{GeorgeetAl08NL,ChoietAl09APL,Karasawa11JITW,Otsuji11SPIE,Strait11NL,Boubanga-TombetetAl12PRB,DochertyetAl12NC,Tielrooij13NatP}, ultrafast THz photocurrent measurements~\cite{PrechteletAl12NC}, and THz-pump and optical probe experiment~\cite{TanietAl12PRL}. Also, negative conductivity in the near-infrared region has been reported~\cite{LietAl12PRL}. When describing carrier relaxation dynamics in graphene, both Coulomb-mediated carrier-carrier and electron-phonon scattering must be taken into account \cite{Rana07PRB,Butscher07APL,Bistritzer09PRL,Romanets10PRB,Winzer10NL,Kim11PRB,Malic11PRB,Sun12PRB,Winzer12PRB,Satou13JAP,Winzer13PRB,Sun13NJP}. Electron-electron interactions such as Auger recombination and impact ionization, which are usually suppressed in conventional semiconductors, play an important role in the relaxation of photo-excited carriers in graphene. Theoretically, Auger recombination is prohibited in clean graphene~\cite{FosterAleiner09PRB} but is expected to become efficient in disordered graphene, which would lead to ultrafast recombination and prevent population inversion. On the other hand, impact ionization is predicted to lead to carrier multiplication \cite{Winzer10NL}, therefore enhancing the efficacy of graphene-based photodetectors. One critical question is whether optically created electrons and holes would thermalize {\em before} recombination and develop separate quasi-Fermi energies. Sun {\it et al}.~\cite{Sun08PRL,Sun11SSC} and Tan {\it et al}.~\cite{TanietAl12PRL} argue that electron-electron interactions are so fast that the system attains a common Fermi energy very quickly. Breusing {\it et al} \cite{Breusing09PRL,Breusing11PRB} and Li {\it et al}. \cite{LietAl12PRL} claim the opposite.



\subsection{THz Nonlinearities}

Mikhailov and Ziegler, using a semiclassical approximation, have shown that  the dynamics of graphene in an AC electric field are intrinsically nonlinear, and efficient frequency multiplications for THz generation can be expected for microwave-driven graphene~\cite{Mikhailov07EPL,MikhailovZiegler08JPCM}.

\begin{figure}
\begin{center}
\includegraphics*[width=0.5\textwidth]{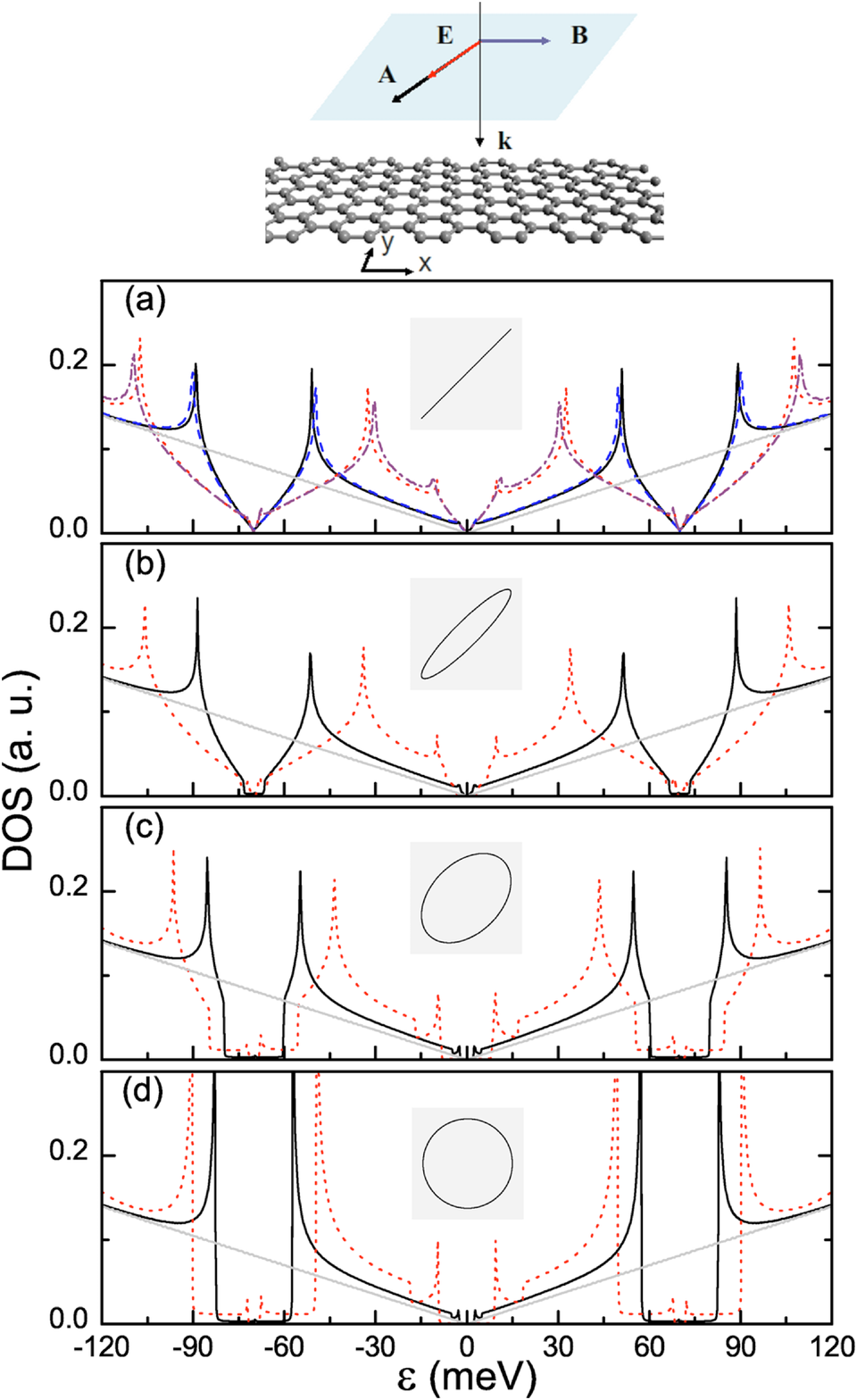}
\end{center}
\caption{
Scheme of the considered setup, where a laser field with $\hbar\Omega=140$ meV ($\Omega$ is the laser frequency) and polarization $\varphi$ is applied perpendicular to a graphene monolayer. DOS for (a) linear, (b) $\varphi=0.125\pi$, (c) $\varphi=0.375\pi$, and (d) circular polarizations. The black solid line is for $I=3.2\times10^{6}\,\mathrm{W}/\ensuremath{\mathrm{cm}}^{2}$ while the red dashed line corresponds
to $I=1.3\times10^{7}\,\mathrm{W}/\ensuremath{\mathrm{cm}}^{2}$. For linear polarization these results are compared
with those of a tight-binding calculation for a system with $5\times10^{4}$
channels (blue dashed and purple dash-dotted lines). For reference the zero-field DOS is shown in solid gray. From~\cite{CalvoetAl11APL}.
}
\label{fig:Calvo}
\end{figure}

Recent theoretical studies also suggest that graphene is an excellent material system for studying solid-state strong-field physics with various promising device possibilities.  One of the predictions is that, when graphene is non-resonantly irradiated by a circularly-polarized laser field, an energy gap should appear at the Dirac point, whose magnitude is proportional to the laser intensity and inversely proportional to the frequency of the laser field \cite{SyzranovetAl08PRB,Lopez-RodriguezNaumis08PRB,OkaAoki09PRB,CalvoetAl11APL,GumbsetAl11JMO,ZhouWu11PRB}. This has tremendous device implications, since many optoelectronic applications require graphene to have a finite band gap, and this prediction provides a coherent and ultrafast means for tuning the band gap.   For example, Calvo and co-workers predicted that a long-wavelength (mid-infrared and THz) laser field can produce significant band gaps in graphene~\cite{CalvoetAl11APL} which can be modulated by the polarization of the field. Figure~\ref{fig:Calvo} shows how the dc density of states (DOS) varies with polarization. For a circularly-polarized mid-infrared laser field at a wavelength of 9\,$\mu$m, the predicted size of the laser-induced gap is 23\,meV for an intensity of 3.2 $\times$ 10$^6$ W/cm$^2$ (which is easily achievable) and increases linearly with the intensity and the inverse of the laser frequency. Thus, the gaps are easily observable with reasonable intensities, and longer wavelengths (such as THz) are preferable.

Nonlinearities are also expected to appear in the dynamics of Dirac fermions in a magnetic field both in the semiclassical regime~\cite{Mikhailov09PRB} and quantum regime~\cite{YaoBelyanin12PRL}.  Mikhailov predicts that due to the linear energy dispersion, the particle responds not only at the resonance frequency but generates a broad frequency spectrum around it. The linewidth of the cyclotron resonance is predicted to be very broad even in a perfectly pure material.  A recent theoretical study by Yao and Belyanin~\cite{YaoBelyanin12PRL} predicts that graphene possesses highest third-order nonlinear optical susceptibility $\chi^{(3)}$. These authors showed that Landau-quantized graphene (i.e., graphene in a perpendicular magnetic field) possesses an extremely high value of $\chi^{(3)}$, of the order of 10$^{-2}$ esu in the mid/far-infrared range in a magnetic field of several tesla. This value is larger than reported $\chi^{(3)}$ values for any known materials by several orders of magnitude, and is a consequence of graphene's unique electronic properties and selection rules near the Dirac point.

Experimentally, THz nonlinear optics in graphene is totally an open field.  Only very recently, Tani and co-workers performed THz-pump optical-probe experiments to demonstrate ultrafast carrier-carrier scattering~\cite{TanietAl12PRL}.

\subsection{THz Plasmonics}



Graphene with its excellent electronic and optical properties makes it a promising candidate for the basis of future optoelectronic devices. Graphene based p-n junctions can be used to separate photo-excited charge carriers. However, the generation of photo-excited carriers is limited to the narrow region of the p-n junction and freestanding graphene only absorbs up to 2.3 $\%$ of incident light \cite{NairetAl08Science,MaketAl08PRL,Kuzmenko08PRL} these factors are hurdles to creating graphene-based devices with high efficiencies. However, absorbtion can be enhanced considerably using plasmonic nanostructures (up to 100\% \cite{Thongrattanasiri12PRL,fang2012graphene}) to radiatively couple to surface plasmon modes, therefore improving the efficiency of graphene-based photodetectors. Furthermore, plasmonic nanostructures can be engineered to be resonant at particular wavelengths therefore opening the door to a new class of wavelength and polarization sensitive graphene-based devices.

Developing technologies to manipulate THz waves is an important goal in THz research.  Several groups have already demonstrated that gated graphene acts as a modulator for THz waves~\cite{HorngetAl11PRB,MaengetAl12NL,RenetAl12NL2,Sensale-RodriguezetAl12NL,Sensale-RodriguezetAl12NC}.  Also, electro-optical modulations in graphene have been considered theoretically by Vasko {\it et al}.~\cite{StrikhaVasko10PRB,StrikhaVasko11JAP,VaskoetAl12PRB}. For both technological and scientific reasons, plasmons in graphene are attracting much attention.  Their unique properties have been intensively studied both theoretically ~\cite{Vafek97PRL,Apalkov07IJMP,Wang07PRB,HwangDasSarma07PRB,RyzhiietAl07JAP2,MikhailovZiegler07PRL,Hanson08JAP,Jablan09PRB,Nikitin11PRB,Koppens11NL,Huidobro12PRB,Peres12JPCM,Diaz12JAP,Davoyan12PRL,Bludov12PRB,Slipchenko13arXiv,GarciaPomar13ACSN,Bludov13IJMP}
and experimentally \cite{JuetAl11NN,Chen12Nat,Fei12Nat,YanNat13Photon,Petkovic13PRL,Gao13NanoLetts}. Many groups have started using gated graphene as well as graphene combined with metallic structures to manipulate THz waves~\cite{JuetAl11NN,CrasseeetAl12NL,YanetAl12NN,LeeetAl12NM}.

A surface plasmon is a coherent fluctuation of charge density in a conducting medium that is restricted to the interface formed between two materials, whose real part of their permittivities differ in sign. Optical excitation of this mode requires that one must provide both the correct energy and in-plane momentum. However, in general, the in-plane momentum of the surface plasmon exceeds that of the photon, and therefore, plane-wave radiation cannot couple to a flat surface. However, momentum matching can be achieved by several schemes, such as prism-coupling using attenuated total internal reflection
\cite{Bludov10EPL,Bludov13IJMP} and by grating coupling \cite{Echtermeyer11NatComs,Zhan12PRB,Gao12ACSN,Peres12Arxiv,Bludov13IJMP,Slipchenko13arXiv},
where the periodicity of the grating provides the additional in-plane momentum needed to excite the surface plasmon. An alternative approach is to spatially modulate graphene's conductivity. Graphene's charge carrier density and hence optical conductivity can be modified by external electric fields and strain, i.e., corrugation. Modulating the optical conductivity in a periodic manner 
allows the optical coupling to surface plasmons \cite{Davoyan12PRL,Peres12JPCM,Bludov12PRB,Bludov13IJMP,Slipchenko13arXiv}. The periodicity of patterned graphene structures such as arrays of nanoribbons \cite{JuetAl11NN,Nikitin12PRB,Bludov13IJMP} and graphene disks \cite{YanetAl12NN,Thongrattanasiri12PRL} can also be used.


Rather unusually, graphene can support both transverse magnetic and transverse electric~\cite{MikhailovZiegler07PRL} plasmon modes in the THz regime, the latter is absent in conventional 2D systems with parabolic electron dispersion. Graphene nanoribbons support both edge and waveguide THz surface plasmon modes, the number of which is a function of ribbon width and frequency
\cite{Nikitin11PRB}. Graphene plasmons can be tuned by electronic gating
\cite{MikhailovZiegler07PRL,Mishchenko10PRL,JuetAl11NN,Thongrattanasiri12APL} as well as by changing the surrounding dielectric environment. The plasmon lifetimes can be extended by increasing the doping level \cite{Jablan09PRB}, and plasmons can be guided by p-n junctions \cite{Mishchenko10PRL}. These properties have lead to several novel device proposals \cite{Popov10PRB,Vakil11Science,Koppens11NL,Echtermeyer11NatComs,JuetAl11NN,Grigorenko12NatPhoton,Chamanara13OptExpress},
ranging from waveguides \cite{Christensen12ACSNano},
switches \cite{Bludov10EPL,LeeetAl12NM,GomezDiaz13arXiv,Amin13SciRep}, and other control devices \cite{Liu11N,Sensale-RodriguezetAl12NC,Chen13IEEETAP} through to sensors \cite{Amin13SciRep}, tunable filters \cite{Serrano13arXiv,Fallahi12PRB,Carrasco13APL} and generators of THz radiation~\cite{RanaIEEE08NanoT,RyzhiietAl09APE,RyzhiietAl08JPCM,Dubinov11JAP,PopovetAl12PRB}.

The first proposals using graphene for antenna applications use graphene as a control element \cite{Dragoman10JAP}.
It was shown that by modulating graphene's conductivity via electronic gating allows the radiation pattern of the deposited metallic dipole antennas to be controlled. A beam reconfigurable antenna was also designed \cite{Huang12IEEENT}, using switchable high impedance surfaces. Several tunable graphene-based antenna schemes using plasmon modes have been developed \cite{Llatser11AIP,Llatser12,Tamagnone12APL,Huidobro12PRB,Filter13OptEx}. Such plasmonic devices offer the possibility of tunable antenna which can be miniaturized down to the nanometer scale, and the recent development of Leaky-Wave Antenna has shown that fixed-frequency beamscanning is also realizable \cite{EsquiusMorote13arXiv}.

\subsection{Detectors}

\begin{figure}
\begin{center}
\includegraphics*[width=0.8\textwidth]{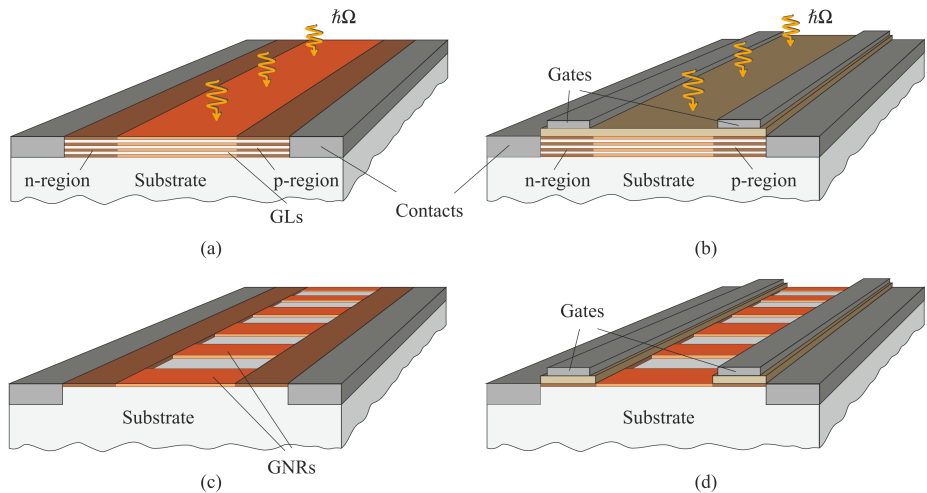}
\end{center}
\caption{
Schematic views of multi-layer graphene p-i-n photodiodes with (a) chemically doped p- and n-regions and (b) electrically induced p- and n-regions (electrical doping), as well as of graphene nanoribbons p-i-n photodiodes with (c) chemically doped p- and n-regions and (d) electrically induced p-and n-regions, respectively. From \cite{Ryzhii12OptRev}.
}
\label{fig:pattern}
\end{figure}

Several schemes using graphene-based heterostructures as THz detectors have been proposed, ranging from devices utilizing randomly stacked multi-layer graphene \cite{Ryzhii09JAP,Ryzhii11IPT,Ryzhii11JJAP,Ryzhii12OptRev}, bilayer graphene \cite{Ryzhii09PRB}, as well as devices based upon arrays of graphene nanoribbons \cite{RyzhiietAl08APE,Ryzhii12IPT,Ryzhii12OptRev,Ryzhii12JOPD} and graphene-based heterostructures which make use of plasmonic effects \cite{RyzhiietAl08JAP,Ryzhii12JOPD}. Some of the proposed devices are shown in figure~\ref{fig:pattern}.


Experimentally, several graphene single- and few-layer photodetectors have been developed, which can operate at room temperature \cite{Xia09NatNanoT,Mueller09NatPhoton,Vicarelli12NatM,Cai13arXiv}. In such devices, the generation of photocurrent is attributed to charge separation due to internal fields, which arise at the graphene-metal interfaces \cite{Lee08NatNano,Mueller09PRB,Xia09NanoLett} and the photo-thermoelectric effect. Recently, broadband (0.76 to 33 THz) graphene transistor detectors have been achieved \cite{kawano2013wide}, which can be tuned by the strength of the applied magnetic field. Plasmonic nanostructures have also been shown to enhance the efficacy of photodetectors and allow such devices to be wavelength and polarization sensitive \cite{Echtermeyer11NatComs}.

Hot electrons have also been utilized in bolometric THz devices operating at cryogenic temperatures made from graphene \cite{Fong12PRX,McKitterick13JAP,McKitterick13arxiv} and bi-layer graphene \cite{Yan12NatNanoT,Kim13PRL}. Bi-layer graphene is a gapless semiconductor with parabolic dispersion \cite{McCann06PRL}, but a band gap can be opened between the conduction and valence bands by applying a potential across the two layers. The band gap and the temperature-dependence of the resistivity is tunable by the strength of the applied field. These properties have been exploited to create bolometric devices that operate at THz frequencies \cite{Yan12NatNanoT,Kim13PRL}. Other bolometric devices use Johnson-noise thermometry \cite{Fong12PRX,McKitterick13JAP,McKitterick13arxiv}, which have sufficient energy resolutions to detect individual THz photons. Room temperature bolometers based on graphene-graphene nanoribbon hybrid heterostructures have also been theoretically proposed \cite{Ryzhii13JOPD}.

The optically induced breakdown of quantum Hall effect has been utilized to create detectors which operate at liquid nitrogen temperatures \cite{Kalugin11APL}, and reduced graphene oxide and graphene nanoribbons infrared photodetectors have also been demonstrated \cite{Chitara11AdvM}. Karch and co-workers reported the observation of the Dynamic Hall Effect and Chiral Edge Currents in THz driven graphene~\cite{KarchetAl10PRL,KarchetAl11PRL} as well as the magnetic quantum ratchet effect \cite{drexler2013magnetic}, and room temperature THz detectors have also been achieved using antenna-coupled graphene based devices \cite{Vicarelli12NatM,Mittendorff13PRL}.

\section{Future Perspectives}

Nanocarbon THz technology has great potential for development in the coming years. Graphene and carbon nanotubes could serve as the basis for new and efficient THz sources and detectors, which are tunable, compact and operate at room temperature. Their desirable electronic and optical properties make them ideally suited for THz antenna and polarizer applications, as well holding the promise of ballistic THz transistors that could supersede traditional silicon technology.

\ack
R.R.H acknowledges financial support from URCO (through Grant No. 17 N 1TAY12-1TAY13). J.K. acknowledges support from the National Science Foundation (through Grants No. OISE-0530220 and EEC-0540832), Department of Energy BES Program (through Grant No. DE-FG02-06ER46308), and the Robert A. Welch Foundation (through Grant No. C-1509). M.E.P was supported by the EU FP7 ITN NOTEDEV (through Grant No. FP7-607521) and FP7 IRSES projects QOCaN (through Grant No. FP7-316432), CANTOR (through Grant No. FP7-612285), and InterNoM (through Grant No. FP7-612624).

\section*{References}
\bibliography{References}

\end{document}